\documentclass[aps,superscriptaddress,prc,twocolumn,nofootinbib,showkeys]{revtex4}
\usepackage{CJK}

\usepackage{graphicx}
\usepackage{epstopdf}
\usepackage{subfigure}
\usepackage{epsfig}
\usepackage{amsmath,amssymb,amsfonts}
\usepackage{color}
\usepackage{xcolor}
\usepackage{url}
\usepackage[utf8]{inputenc}
\usepackage[bookmarks,
                bookmarksopen = true,
                bookmarksnumbered = true,
                linktocpage,
                colorlinks = true,
                linkcolor = blue,
                urlcolor  = blue,
                citecolor = blue,
                anchorcolor = green,
                hyperindex = true,
                hyperfigures]
                {hyperref}

\begin{document}

\begin{CJK*}{UTF8}{gbsn} 
\title{Effects of the formation time of parton shower on jet quenching in heavy-ion collisions\footnote{Supported by the National Natural Science Foundation of China under Grant Nos.~12175122, 2021-867, 11890710, 11890713, and 14-547.}
}

\author{Mengxue Zhang ({\CJKfamily{gbsn}张梦雪})}
\affiliation{Institute of Frontier and Interdisciplinary Science, Shandong University, Qingdao, Shandong 266237, China}\affiliation{Key Laboratory of Particle Physics and Particle Irradiation of Ministry of Education, Shandong University, Qingdao, Shandong, 266237, China}

\author{Yang He ({\CJKfamily{gbsn}何杨})}
\affiliation{Institute of Frontier and Interdisciplinary Science, Shandong University, Qingdao, Shandong 266237, China}\affiliation{Key Laboratory of Particle Physics and Particle Irradiation of Ministry of Education, Shandong University, Qingdao, Shandong, 266237, China}

\author{Shanshan Cao ({\CJKfamily{gbsn}曹杉杉})}
\email{shanshan.cao@sdu.edu.cn}
\affiliation{Institute of Frontier and Interdisciplinary Science, Shandong University, Qingdao, Shandong 266237, China}\affiliation{Key Laboratory of Particle Physics and Particle Irradiation of Ministry of Education, Shandong University, Qingdao, Shandong, 266237, China}

\author{Li Yi ({\CJKfamily{gbsn}易立})}
\email{li.yi@sdu.edu.cn}
\affiliation{Institute of Frontier and Interdisciplinary Science, Shandong University, Qingdao, Shandong 266237, China}\affiliation{Key Laboratory of Particle Physics and Particle Irradiation of Ministry of Education, Shandong University, Qingdao, Shandong, 266237, China}

\date{\today}


\begin{abstract}

Jet quenching has successfully served as a hard probe to study the properties of Quark-Gluon Plasma (QGP). As a multi-particle system, jets take time to develop from a highly virtual parton to a group of partons close to mass shells. In this study, we present a systematical study on the effects of this formation time on jet quenching in relativistic nuclear collisions. Jets from initial hard scatterings were simulated with Pythia, and their interactions with  QGP were described using a Linear Boltzmann Transport (LBT) model that incorporates both elastic and inelastic scatterings between jet partons and the thermal medium. Three different estimations of the jet formation time were implemented and compared, including instantaneous formation, formation from single splitting, and formation from sequential splittings, before which no jet-medium interaction was assumed. We found that deferring the jet-medium interaction with a longer formation time not only affects the overall magnitude of the nuclear modification factor of jets, but also its dependence on the jet transverse momentum.

\end{abstract}


\maketitle
\end{CJK*}


\section{Introduction}
\label{sec:introduction}

Quark-gluon plasma (QGP) is a state of matter in which quarks and gluons are deconfined instead of being bounded inside hadrons~\cite{Gyulassy:2004zy,Jacobs:2004qv,Busza:2018rrf}. Relativistic heavy-ion collisions provide a unique laboratory to study the properties of QGP~\cite{Busza:2018rrf}, and jet quenching is among the major signatures of the creation of QGP in these energetic collisions~\cite{Qin:2015srf,Cao:2020wlm,Connors:2017ptx}. The observed suppression of the high transverse momentum ($p_\mathrm{T}$) hadron and reconstructed jet spectra is considered a consequence of both elastic and inelastic scatterings between the energetic partons produced via initial hard collisions and the color-deconfined QGP medium~\cite{Wang:1991xy,Gyulassy:1993hr,Zakharov:1997uu,Baier:1998yf,Wang:2001ifa,Arnold:2002ja,Gyulassy:2003mc,Kovner:2003zj,Qin:2007rn,Bass:2008rv,Majumder:2009ge,Armesto:2011ht,Zhang:2019toi,Sirimanna:2021sqx}. The amount of energy transferred between jet partons and  QGP is governed by a set of transport coefficients, such as the strong coupling parameter $\alpha_\mathrm{s}$ and jet quenching parameter $\hat{q}$~\cite{Baier:2002tc,Majumder:2008zg}. And it is still an ongoing effort to extract these parameters from the jet quenching data, such as the nuclear modification factor~\cite{JET:2013cls,JETSCAPE:2021ehl,Xie:2022ght}, which helps quantify the transport properties or opacity of the QGP medium. 

With tremendous efforts on systematical experimental measurements and ever more sophisticated theoretical calculations, studies on jet-medium interactions have been extended from nuclear modification of high $p_\mathrm{T}$ hadrons~\cite{Vitev:2002pf,Salgado:2003gb,Dainese:2004te,Wicks:2005gt,Armesto:2005iq,Chen:2011vt,Cao:2017hhk,Xing:2019xae} and jets~\cite{Aad:2014bxa, Khachatryan:2016jfl,Qin:2010mn,Young:2011qx,Dai:2012am,Wang:2013cia,Blaizot:2013hx,Mehtar-Tani:2014yea,Cao:2017qpx,Kang:2017frl,He:2018xjv,He:2022evt,JETSCAPE:2022jer}, to the intra-structures of jets~\cite{Ramos:2014mba,Lokhtin:2014vda,Chien:2015hda,Casalderrey-Solana:2016jvj,Tachibana:2017syd,KunnawalkamElayavalli:2017hxo,Park:2018acg,Luo:2018pto,Chang:2017gkt,Mehtar-Tani:2016aco,Milhano:2017nzm,Caucal:2019uvr,Chen:2020tbl} as well as jet-related correlations~\cite{Aad:2010bu,Chatrchyan:2012gt,Qin:2009bk,Chen:2016vem,Chen:2016cof,Chen:2017zte,Zhang:2018urd,Kang:2018wrs,Yang:2021qtl,Luo:2021voy}. In most of these studies, jet production in heavy-ion collisions is usually divided into three stages: parton production and shower in vacuum (or proton-proton collisions), interaction with QGP, and hadronization. However, different assumptions have been adopted for the starting time of jet-medium interactions. This could introduce uncertainties in evaluating the nuclear modification on jets and has attracted several investigations in recent literature~\cite{Apolinario:2020uvt,Adhya:2021kws}. For instance, this starting time has shown to affect the azimuthal dependence of jet quenching~\cite{Adhya:2021kws}. Moreover, with a time reclustering algorithm, jets with longer formation time exhibit a weaker quenching~\cite{Apolinario:2020uvt}. 

In this study, we will conduct a detailed study on the effects of the jet formation time on jet energy loss. The initial jets prior to the interaction with QGP were generated from Pythia 8 simulation~\cite{Sjostrand:2014zea,Sjostrand:2006za}. Due to the lack of  information on the jet formation time from Pythia, we designed three different evaluations on the production time of each parton within jets, varying from zero formation time, to an estimation based on a single splitting before formation, and a more elaborate estimation for a sequence of multiple splittings generated in Pythia. Interactions between these jet partons and QGP were then simulated using a linear Boltzmann transport (LBT) model~\cite{Cao:2016gvr,Cao:2017hhk} that describes both elastic and inelastic scatterings between jet partons and thermal partons from QGP. Within this framework, we investigated how different estimations of the jet parton formation time affects the nuclear modification of fully reconstructed jets at the RHIC energy, and found that different modelings of this formation time impact not only the overall magnitude but also the $p_\mathrm{T}$ dependence of the nuclear modification factor of jets. The goal of the present study was to explore the sensitivity of jet quenching to the starting time of parton-medium interactions. For a more comprehensive discussion on the LBT model and its comparison with various experimental data, one may refer to Refs.~\cite{Cao:2016gvr,Cao:2017hhk,Chen:2017zte,Luo:2018pto,He:2018xjv,Xing:2019xae,Chen:2020tbl,Yang:2021qtl,He:2022evt,Yang:2022nei}.

The remainder of this paper is organized as follows. In Sec.~\ref{sec:formationTime}, we discuss how the parton shower generated by Pythia was used to estimate the formation time of each parton, and study the dependence of the formation time on the parton energy. In Sec.~\ref{sec:LBT}, we investigate effects of formation time on the nuclear modification factor ($R_\mathrm{AA}$) and central-to-peripheral ratio ($R_\mathrm{cp}$) of jets in heavy-ion collisions using the LBT model. The paper is summarized in Sec.~\ref{sec:summary}.

\section{Modelings of the parton formation time}
\label{sec:formationTime}

We used the Pythia 8 event generator to simulate jet parton production and its vacuum shower. Since initial parton production processes -- e.g., multi-parton interaction (MPI) -- other than the hardest scattering are also essential in describing jet observables, especially those related to soft particles~\cite{Yang:2021qtl,STAR:2019cie}, we fed full Pythia events of final-state partons into the LBT model for their subsequent interactions with QGP. In hard scatterings, a pair of highly virtual partons are first created, which continue splitting until the virtuality of each daughter is sufficiently low -- close to its mass shell or approaching the scale of hadronization. We used the mother-daughter tree provided by the Pythia shower to evaluate the time taken by each splitting, thereby obtaining the formation time of each parton. 

For a $1\rightarrow 2$ process, the splitting time can be estimated using the uncertainty principle as~\cite{Adil:2006ra}
\begin{equation}
\label{eq:formationTime}
\tau_\mathrm{form}=\frac{2Ex(1-x)}{k_\perp^2},
\end{equation}
in which $E$ represents the energy of the mother parton, $x$ and $(1-x)$ are the energy fractions taken by the two daughters, and $k_\perp$ is the transverse momentum of the daughters with respect to their mother. Here, the rest masses of both mother and daughters are neglected. Since $k_\perp^2/[x(1-x)]$ gives the virtuality ($Q^2$) of the mother parton, the formation time can also be written as $\tau_\mathrm{form}=2E/Q^2$. Since the uncertainty principle $\Delta x \Delta p \sim 1$ has been used to obtain Eq.~(\ref{eq:formationTime}), one should treat this relation as an approximation of the same order. Therefore, it is necessary to understand the sensitivity of jet energy loss to the exact values applied for the formation time. In the literature, $\tau_\mathrm{form}\sim2E/k_\perp^2$ is also frequently used by assuming that $k_\perp^2$ and $Q^2$ are of the same order. 

To study the effects of parton formation time on jet quenching, we compared our calculations for three different estimations of this formation time.
\begin{itemize}
\item {\bf Setup 1: zero formation time} -- the vacuum shower is assumed to happen instantaneously ($\tau_1=0$) and jet partons start to interact with QGP when the hydrodynamic evolution of QGP commences (at $\tau_0$).
\item {\bf Setup 2: formation time from single splitting} -- each parton is assumed to be formed from one splitting which takes the time of $\tau_2=2Ex(1-x)/k_\perp^2$, where $E$ represents the energy of the ancestor parton (directly produced by the initial hard scattering) at the top of the mother-daughter tree generated by Pythia shower, $x$ and $k_\perp$ are respectively the fractional energy and transverse momentum of the given final-state parton with respect to its ancestor; thus this jet parton starts to interact with QGP at $t_\mathrm{init}=\max (\tau_0,\tau_2)$.
\item {\bf Setup 3: formation time from multiple splittings} -- the full sequence of splittings from the very first ancestor to each final-state parton in Pythia is tracked, and the parton formation time is calculated as $\tau_3=\sum_i 2E_i x_i(1-x_i)/k_{\perp i}^2$, where $E_i$ represents the energy of the mother parton in the $i$-th splitting, $x_i$ and $k_{\perp i}$ are respectively the fractional energy and transverse momentum of a daughter with respect to the mother; thus this jet parton starts to interact with QGP at $t_\mathrm{init}=\max (\tau_0,\tau_3)$.
\end{itemize}
These setups are well defined for partons originating from the initial hard scattering process. For those from other sources in Pythia simulation, such as the initial state radiation, we set their formation time as zero in the present work. Theoretically, we consider that setup-3 is a better choice than setup-2. However, given that setup-2 is widely applied in the literature as a quick estimation of the hard parton formation time from jet vacuum showers, it is worth investigating their difference on jet observables within the same framework.

\begin{figure}[tbp]
\includegraphics[width=0.85\linewidth]{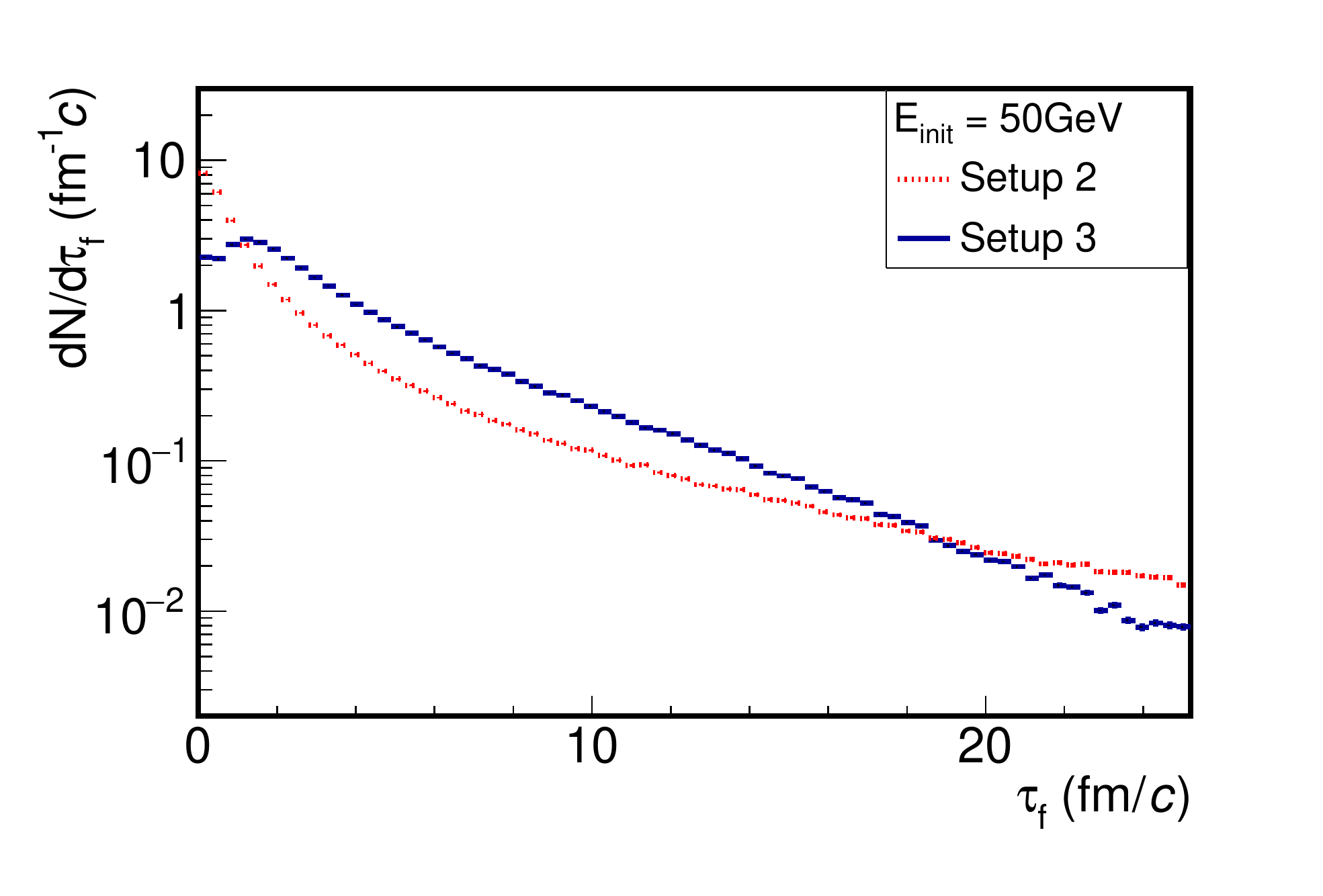}
\caption{(Color online) Distribution of final-state partons from Pythia emanating from a 50~GeV quark as functions of the formation time, compared between different setups.}
\label{fig:ftDistribution}
\end{figure}

In Fig.~\ref{fig:ftDistribution} we first study the formation time distribution of the final-state Pythia partons developed from a single quark at a fixed energy of 50~GeV and maximum possible virtuality scale of 50~GeV, compared between setup-2 (single splitting) and setup-3 (sequential splittings). One would expect a $\delta$-function at zero formation time for setup-1 (instantaneous formation). Compared to setup-1, one observes that a large number of partons from setup-2 and 3 are formed during the QGP phase: there are only approximately 50\% partons for setup-2 and 20\% partons for setup-3 formed before $\sim 1$~fm/$c$ (the scale of the initial time of QGP); the amount increases to about 83\% and 76\% at the time around 5~fm/$c$ and approaches about 90\% and 92\% around 10~fm/$c$ (the QGP lifetime). The remaining amount is formed out of dense nuclear matter. Therefore, taking into account the parton formation time significantly delays the jet-medium interaction and affects the jet quenching observables.

The difference in the formation time between setup-2 and 3 originates from two competing effects. The addition of time for a sequence of splittings (setup-3) can lead to a longer formation time than that of a single splitting (setup-2). On the other hand, since  both energy $E$ and virtuality $Q^2$ (or $k_\perp^2$) in Eq.~(\ref{eq:formationTime}) drop after each splitting, it is possible that the formation time estimated from setup-2 is larger than that from setup-3. In general, the parton distributions are comparable (of the same order) over a wide range of formation time in Fig.~\ref{fig:ftDistribution}. This can be understood with the dominating contribution to the total formation time from the last (softest) splitting. However, a closer comparison suggests that, within the QGP lifetime ($1\sim 10$~fm/$c$), partons from setup-3 tend to form later than those from setup-2.



\begin{figure}[tbp]
\includegraphics[width=0.85\linewidth]{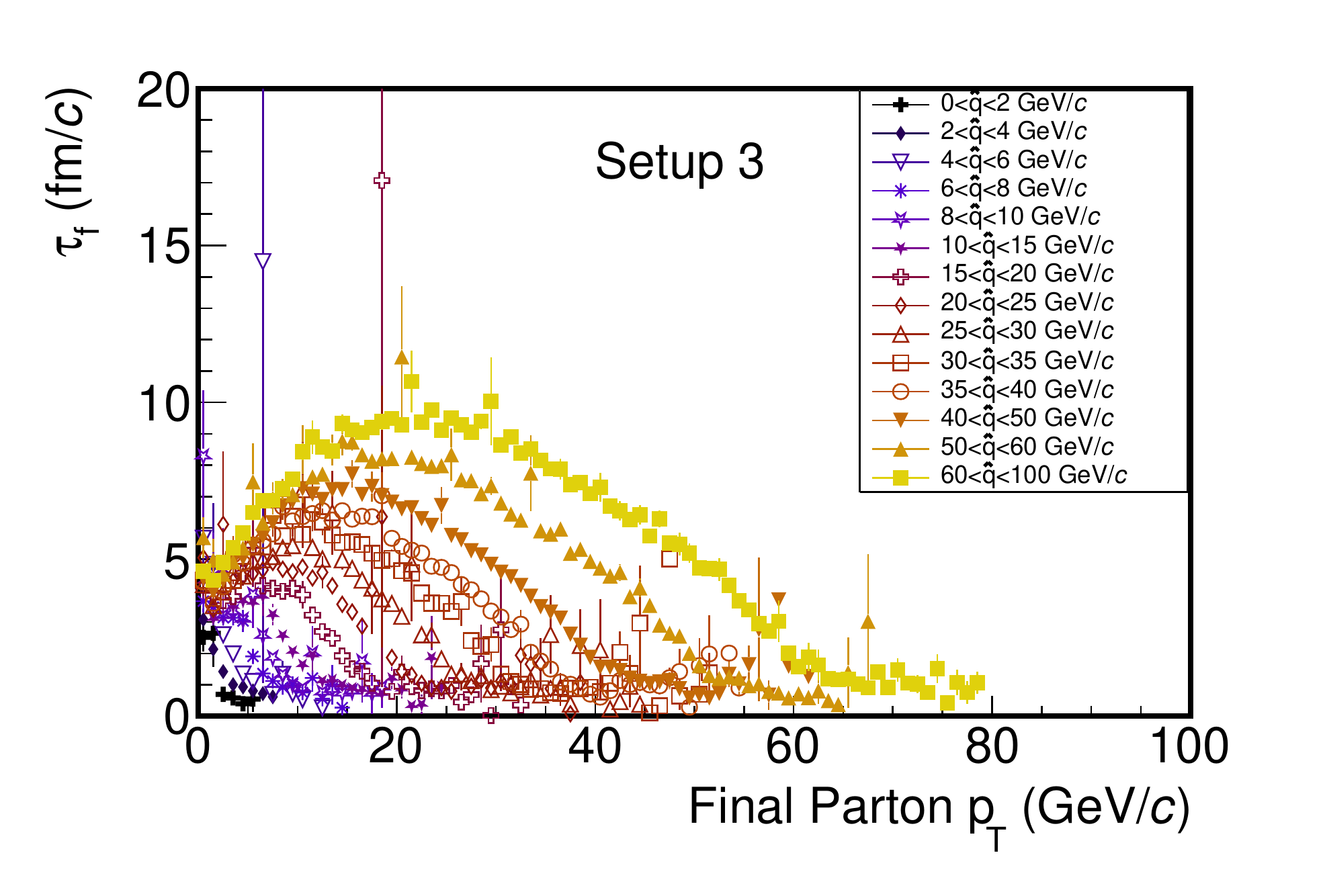}
\caption{(Color online) Dependence of the average formation time from sequential splittings (setup-3) on the final-state parton $p_\mathrm{T}$, compared between different $\hat{q}$ regions in Pythia for p+p collisions at $\sqrt{s}=200$~GeV.}
\label{fig:ftvsFinalpT}
\end{figure}

To further investigate the dependence of the formation time on the parton energy scale, Fig.~\ref{fig:ftvsFinalpT} presents the average formation time of partons as a function of their final state $p_\mathrm{T}$ generated by Pythia. Here, we simulated proton-proton (p+p) collisions $\sqrt{s}=200$~GeV and used setup-3 for sequential splittings to calculate the parton formation time. Inside the figure, results from different $\hat{q}$ bins are compared, which governs the amount of momentum exchange for the initial hard scatterings in Pythia and is around the initial $p_\mathrm{T}$ of partons directly produced from the hard splittings. From Fig.~\ref{fig:ftvsFinalpT}, we observe that for a given $\hat{q}$ bin, the formation time first increases and then decreases as the final-state parton $p_\mathrm{T}$ increases. Since the hardest final-state partons are most likely produced via very few unbalanced splittings (or even no splitting) from the initial hard parton, they exhibit a short formation time. By contrast, medium $p_\mathrm{T}$ partons that approach mass shells after multiple splittings show a longer formation time. We have also noticed that the longest formation time comes from splittings where daughter partons are almost collinear ($k_\perp\rightarrow 0$). The peak value of the formation time becomes larger as one increases the initial $\hat{q}$ bin. This is because partons produced from more energetic collisions usually possess higher virtualities thus take longer time to shower towards their mass shells.

\begin{figure}[tbp]
\includegraphics[width=0.85\linewidth]{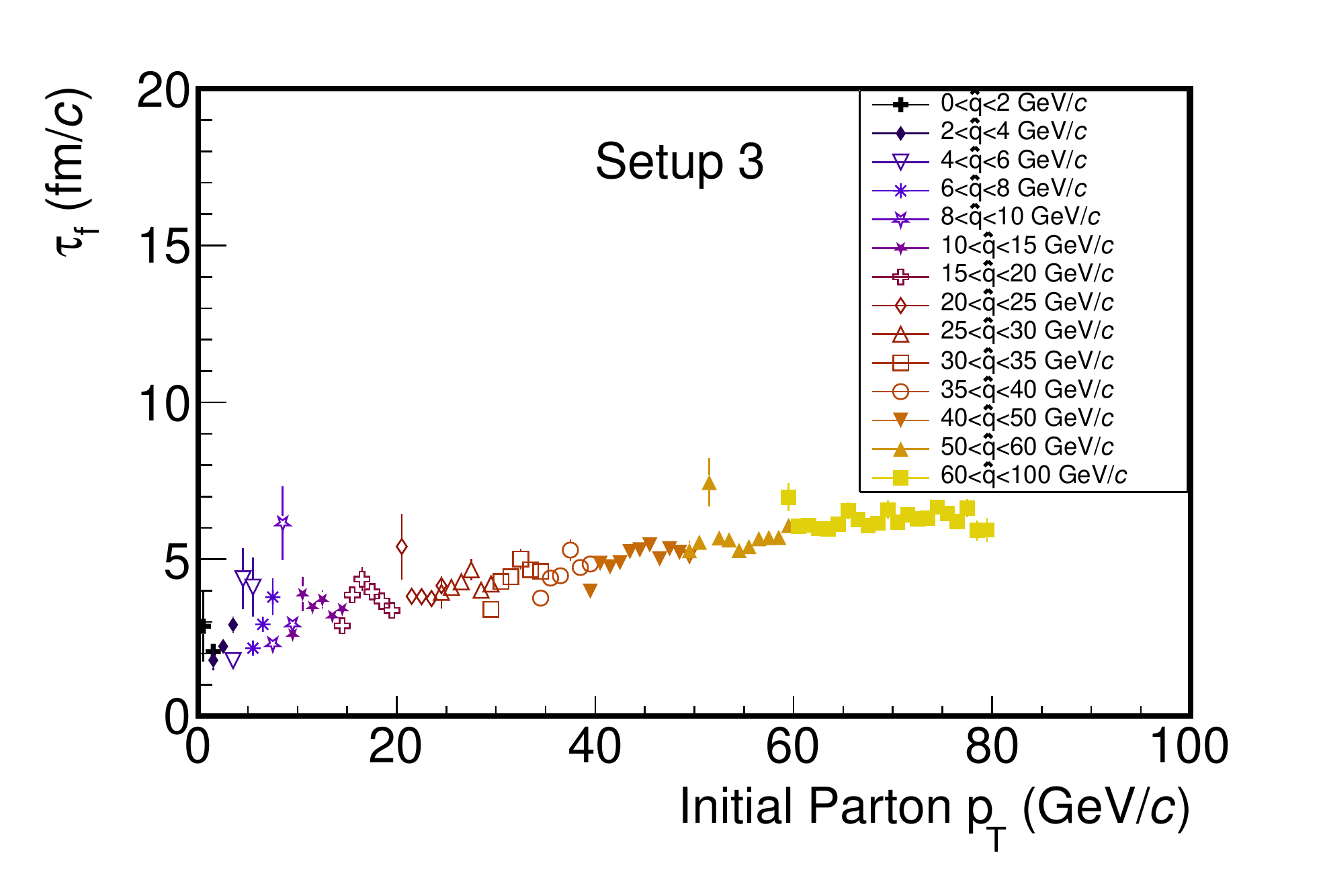}
\caption{(Color online) Dependence of the average formation time from sequential splittings (setup-3) on the initial-state parton $p_\mathrm{T}$, compared between different $\hat{q}$ regions in Pythia for p+p collisions at $\sqrt{s}=200$~GeV.}
\label{fig:ftvsInitialpT}
\end{figure}

The same conclusions can also be drawn from Fig.~\ref{fig:ftvsInitialpT} where we present the average formation time as a function of the $p_\mathrm{T}$ of the ancestor partons directly produced from hard collisions in Pythia. In this figure, one can clearly observe the mapping between the $\hat{q}$ bins and the ranges of the ancestor parton $p_\mathrm{T}$. With the increase of this initial $p_\mathrm{T}$, the average time for shower partons to approach their mass shells becomes longer. An approximately linear relation can be seen between the average formation time of the shower partons and the initial $p_\mathrm{T}$ of the ancestor partons. We confirmed that the parton formation time estimated using setup-2 (single splitting scenario) shares the similar dependences on the initial and final state parton $p_\mathrm{T}$ to setup-3 here.

\section{Nuclear modification of jets}
\label{sec:LBT}

The final-state partons generated by Pythia were fed into the linear Boltzmann transport (LBT) model~\cite{Cao:2016gvr,Cao:2017hhk} for their subsequent interactions with the QGP medium. In LBT, the phase space distribution of jet partons (denoted by ``$a$" here) evolves according to the Boltzmann equation 
\begin{equation}
p_a \cdot \partial f_a= E_a (\mathcal{C}_\mathrm{el}+\mathcal{C}_\mathrm{inel}),
\label{eq:Boltzmann}
\end{equation}
where the collision term on the right hand side includes contributions from both elastic and inelastic processes. From the collision term, one may extract the elastic scattering rate as
\begin{align}
\label{eq:rate}
\Gamma_a^\mathrm{el}&(\vec{p}_a,T)=\sum_{b,(cd)}\frac{\gamma_b}{2E_a}\int \prod_{i=b,c,d}\frac{dp_i^3}{E_i(2\pi)^3} f_b S_2(\hat{s},\hat{t},\hat{u})\nonumber\\
&\times (2\pi)^4\delta^{(4)}(p_a+p_b-p_c-p_d)|\mathcal{M}_{ab\rightarrow cd}|^2,
\end{align}
in which we sum over all possible $ab\rightarrow cd$ channels, $\gamma_b$ and $f_b$ represent the color-spin degrees of freedom and the distribution of thermal partons inside QGP respectively, and the function $S_2(\hat{s},\hat{t},\hat{u})=\theta(\hat{s}\ge 2\mu_\mathrm{D}^2)\theta(-\hat{s}+\mu^2_\mathrm{D}\le \hat{t} \le -\mu_\mathrm{D}^2)$ is introduced~\cite{Auvinen:2009qm} to avoid the collinear divergence in the leading-order (LO) scattering matrices $\mathcal{M}_{ab\rightarrow cd}$, with $\hat{s}$, $\hat{t}$ and $\hat{u}$ denoting the Mandelstam variables and $\mu_\mathrm{D}$ denoting the Debye screening mass. Meanwhile, the inelastic scattering rate is related to the average number of medium-induced gluons per unit time as
\begin{equation}
 \label{eq:gluonnumber}
 \Gamma_a^\mathrm{inel} (E_a,T,t) = \int dxdk_\perp^2 \frac{dN_g^a}{dx dk_\perp^2 dt},
\end{equation}
where the gluon spectrum is taken from the higher-twist energy loss calculation~\cite{Wang:2001ifa,Zhang:2003wk,Majumder:2009ge},
\begin{equation}
\label{eq:gluondistribution}
\frac{dN_g^a}{dx dk_\perp^2 dt}=\frac{2C_A\alpha_\mathrm{s} P^\mathrm{vac}_a(x)}{\pi C_2(a) k_\perp^4}\,\hat{q}_a\, {\sin}^2\left(\frac{t-t_i}{2\tau_f}\right).
\end{equation}
In the above equation, $x$ and $k_\perp$ are the fractional energy and transverse momentum of the emitted gluon with respect to its parent parton, $P^\mathrm{vac}_a(x)$ is the vacuum splitting function of $a$ with the color factor $C_2(a)$ included, $\hat{q}_a$ is the parton transport coefficient that characterizes the transverse momentum broadening square per unit time due to elastic scatterings, $t_i$ represents the production time of parton $a$, and $\tau_f={2E_a x(1-x)}/k_\perp^2$ is taken as the formation time of the emitted gluon in LBT. The formation time (or production time) of jet partons discussed in the previous section will directly determine when they start these elastic and inelastic scatterings with the QGP medium. The only parameter we adjusted for LBT in this study is the strong coupling constant $\alpha_\mathrm{s}$, which directly affects the interaction strength in elastic scatterings, and controls the rate of medium-induced gluon through the jet transport coefficient $\hat{q}_a$.

For realistic heavy-ion collisions, the spatial distribution of initial jets was calculated according to the binary collision vertices from the Monte-Carlo (MC) Glauber model. QGP was simulated using a viscous hydrodynamic model (VISHNew~\cite{Song:2007fn,Song:2007ux,Qiu:2011hf} in this study) whose entropy density distribution was initialized using the MC Glauber model. The initial time of the hydrodynamic evolution was set as $\tau_0=0.6$~fm/$c$, and the specific shear viscosity was set as $\eta/s=0.08$ for a reasonable description of the soft hadron observables at RHIC and LHC. This hydrodynamic model provides the spacetime information of the local temperature and flow velocity of the QGP medium, based on which we obtained the momentum distribution of thermal partons that enters the collision term on the right hand side of Eq.~(\ref{eq:Boltzmann}). 

In the LBT model, we not only tracked the phase space evolution of the jet partons and their emitted gluons, but also the thermal partons that scattered out of the QGP background. The latter is denoted as ``recoil" partons. In addition, generation of these recoil partons leaves particle holes inside the QGP, which are denoted as back-reaction or ``negative" partons, and also tracked inside LBT in order to guarantee the energy-momentum conservation of the parton system. Recoil and negative partons constitute the ``jet-induced medium excitation" and have been shown essential in understanding observables related to fully reconstructed jets~\cite{He:2018xjv,He:2022evt}. At the chemical freezeout hypersurface ($T_\mathrm{c}=165$~MeV), all partons discussed above were collected for jet reconstruction and observable analysis. Their further interaction with the hadron gas was neglected, considering its much more dilute density compared to the QGP medium.

For jet reconstruction, we fed all partons from Pythia (for p+p collisions) or Pythia+LBT (for heavy-ion, or A+A, collisions) into the Fastjet package with the anti-$k_\perp$ algorithm selected~\cite{Cacciari:2005hq,Cacciari:2011ma}. In this study, particles in the mid-rapidity $|\eta|<1$ and $p_\mathrm{T}>0.2$~GeV/$c$ are used for constructing jets. For a given jet cone size $R$, the reconstructed jet $\eta_\mathrm{jet}$ was required to be at $R$ distance away from the acceptance edge as $|\eta_\mathrm{jet}|<1-R$, so that the full jet located inside the acceptance coverage. Note that the energy-momentum of the negative partons produced by LBT is subtracted from the reconstructed jets, similar to subtracting the medium background. 

\begin{figure}[tbp]
\includegraphics[width=0.9\linewidth]{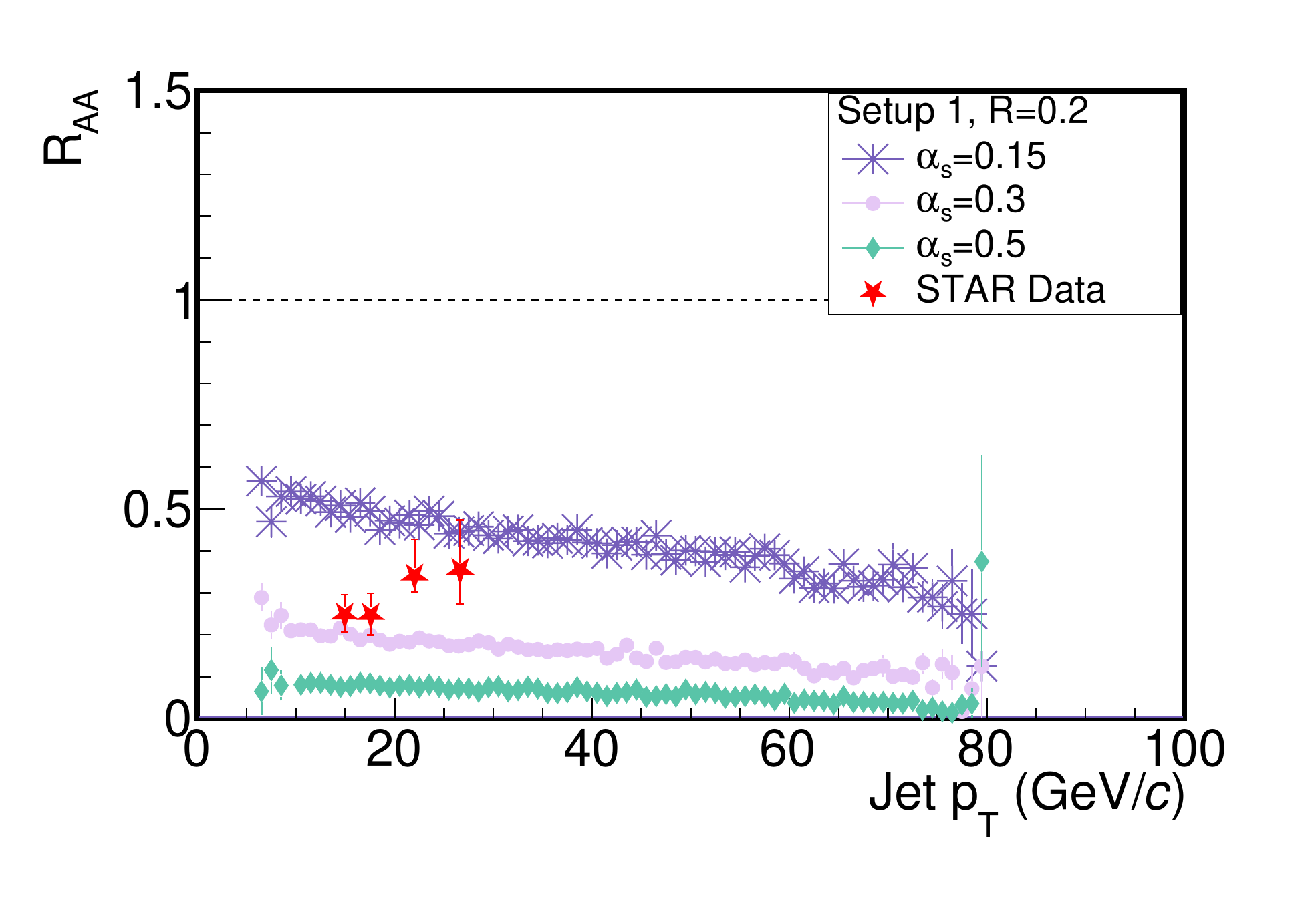}
\includegraphics[width=0.9\linewidth]{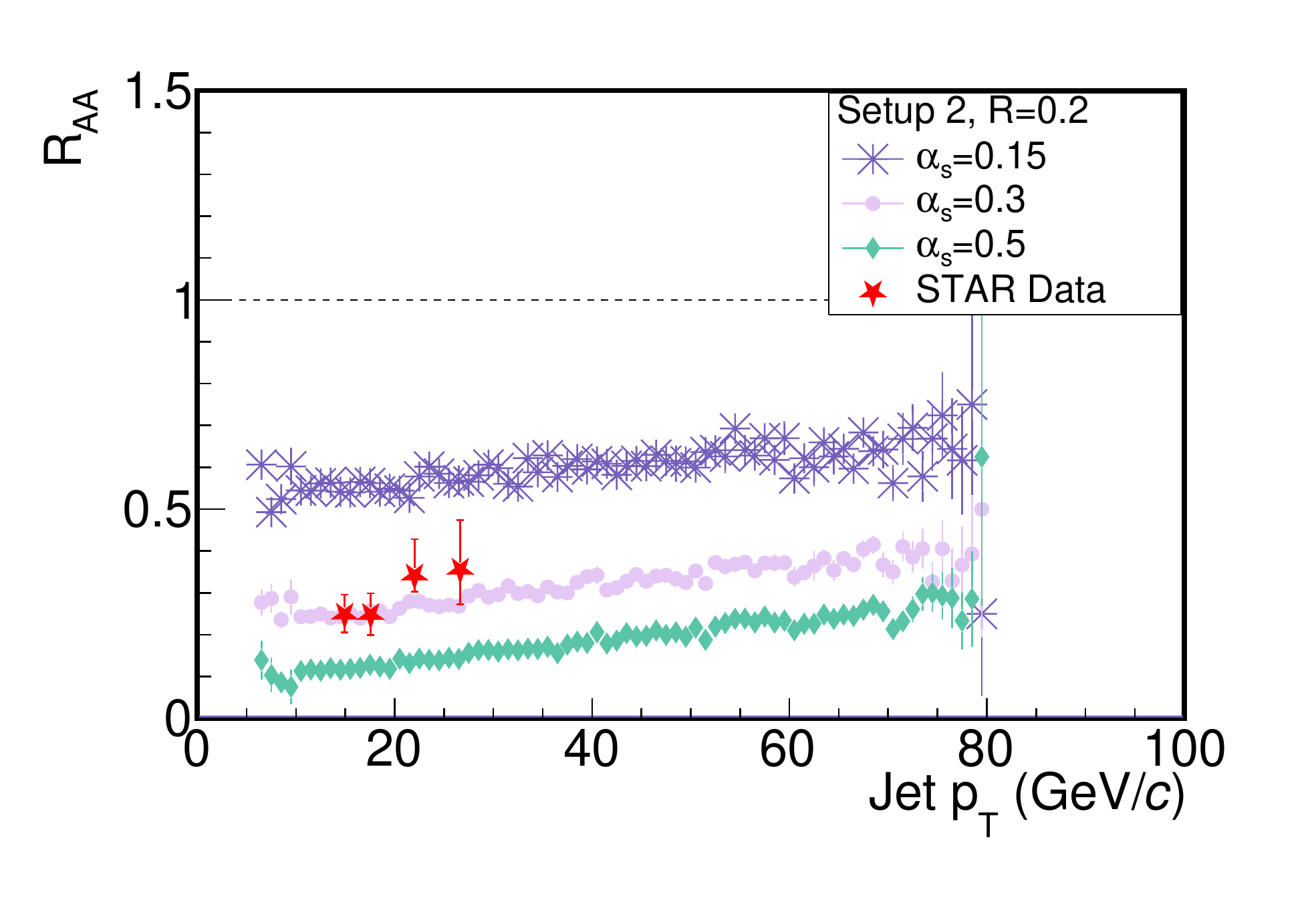}
\includegraphics[width=0.9\linewidth]{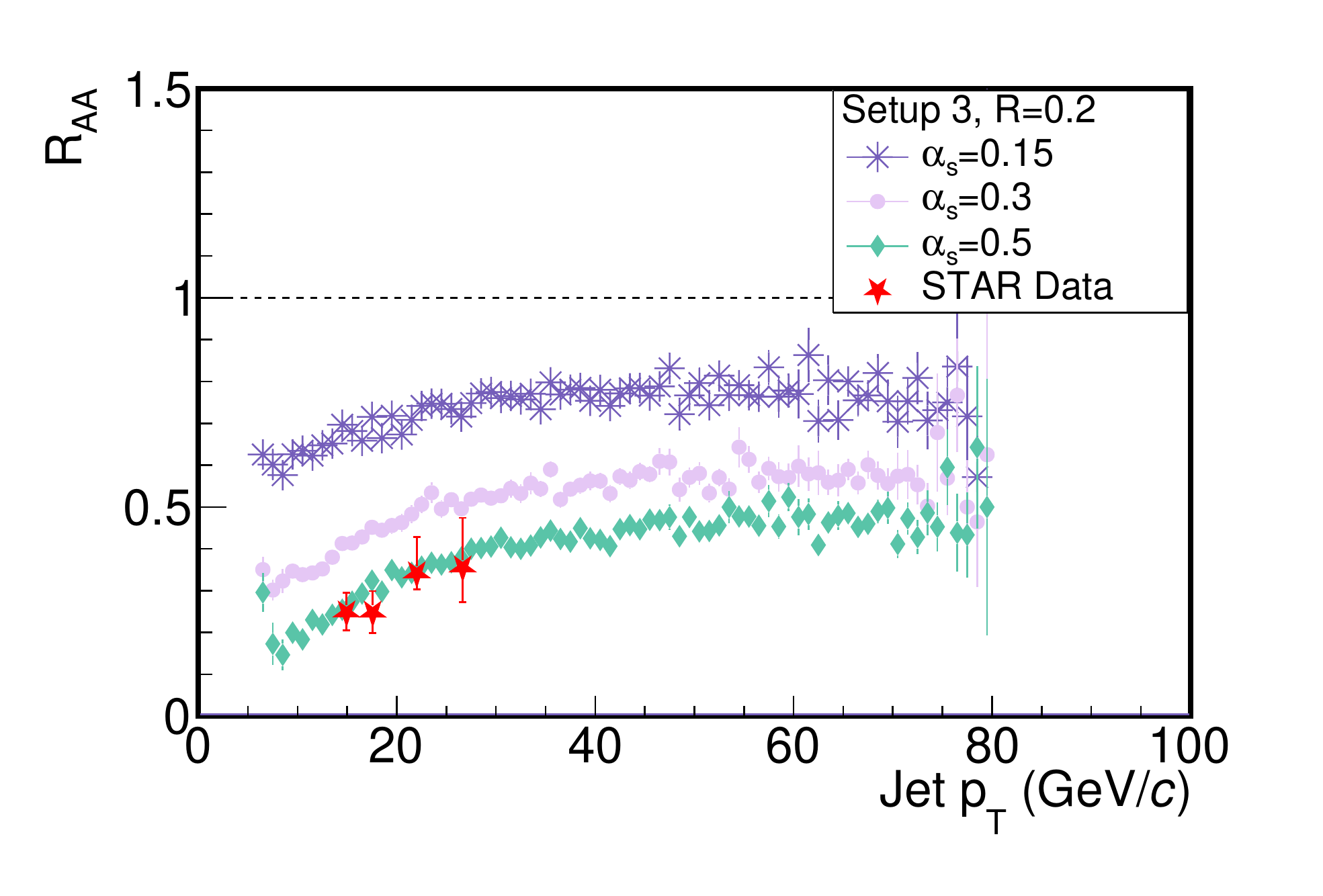}
\caption{(Color online) Nuclear modification factor $R_\mathrm{AA}$ of jets in 0-10\% Au+Au collisions at $\sqrt{s_\mathrm{NN}}=200$~GeV, compared between using different $\alpha_\mathrm{s}$ values and different setups of the parton formation time -- upper panel for setup-1, middle for setup-2 and lower for setup-3.}
\label{fig:Raa}
\end{figure}

\begin{figure}[tbp]
\includegraphics[width=0.9\linewidth]{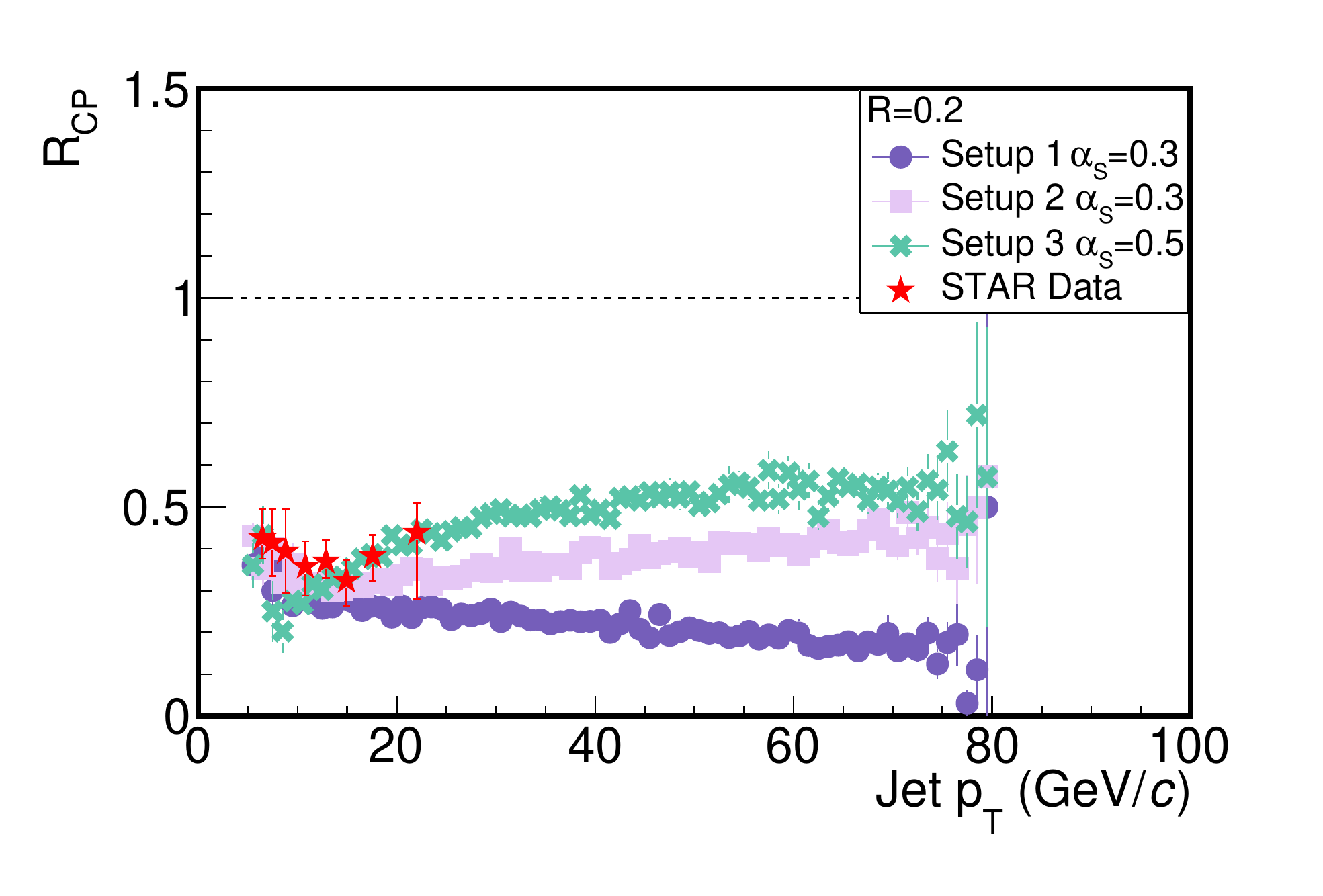}
\includegraphics[width=0.9\linewidth]{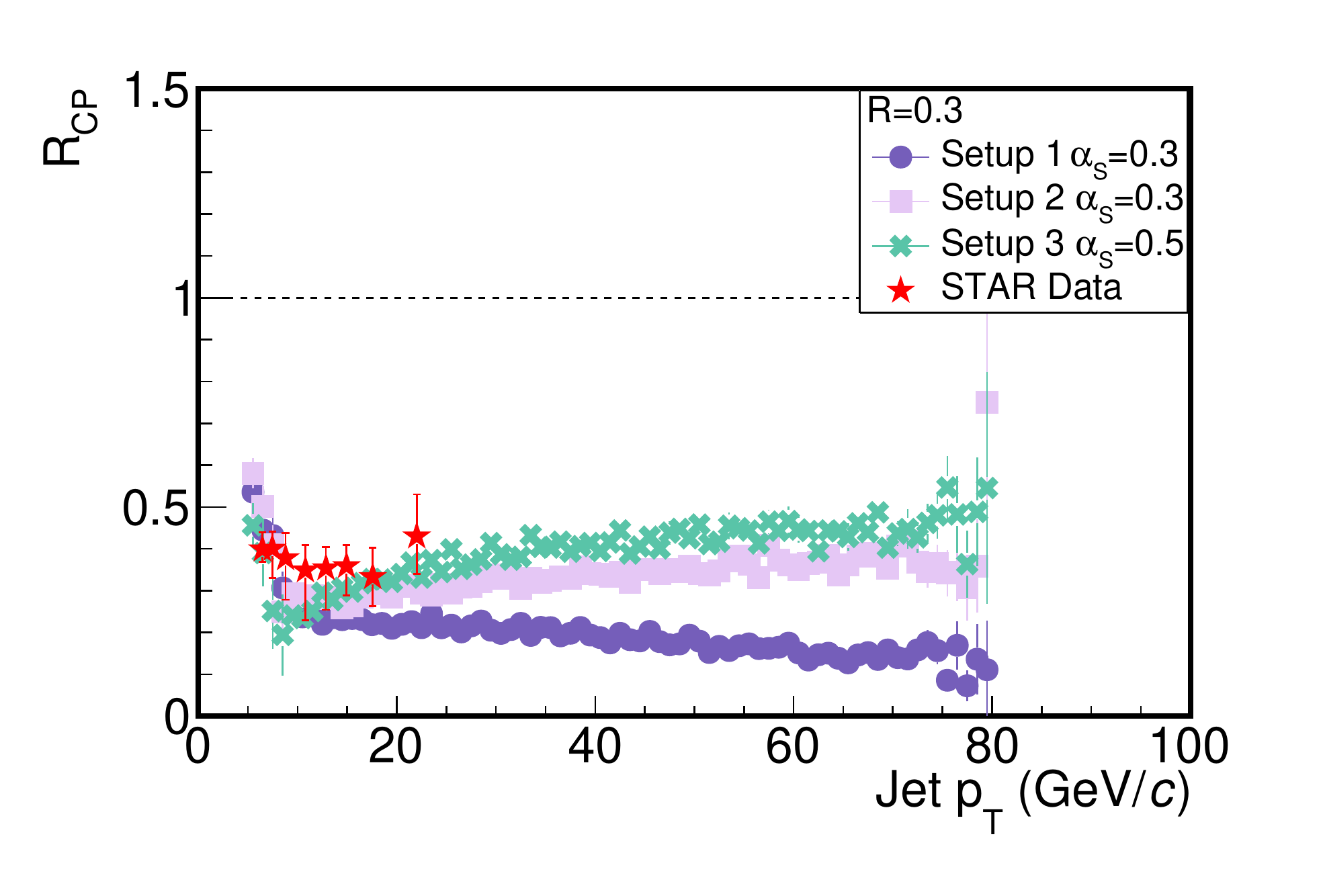}
\includegraphics[width=0.9\linewidth]{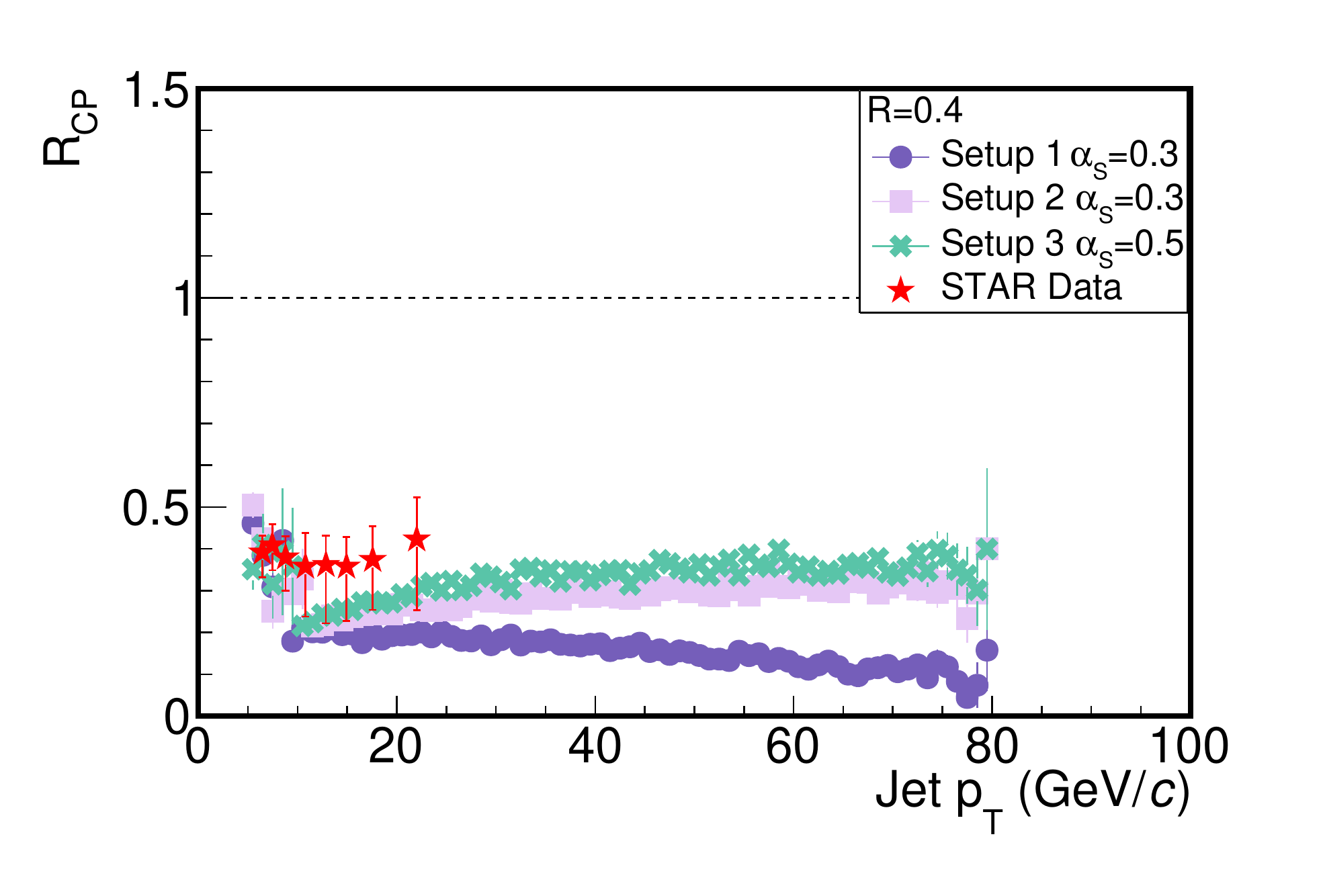}
\caption{(Color online) Central (0-10\%) to peripheral (60-80\%) ratio ($R_\mathrm{cp}$) of jets in Au+Au collisions at $\sqrt{s_\mathrm{NN}}=200$~GeV, compared between different setups of the parton formation time at different jet cone sizes, upper panel for $R=0.2$, middle for $R=0.3$ and lower for $R=0.4$.}
\label{fig:Rcp}
\end{figure}

Shown in Fig.~\ref{fig:Raa} is the nuclear modification factor $R_\mathrm{AA}$ of full jets with a cone size $R=0.2$ in the top 10\% Au-Au collisions at $\sqrt{s_\mathrm{NN}}=200$~GeV. Between different panels, we compare the three proposed setups of parton formation time, upper panel for instantaneous formation (setup-1), middle for single splitting (setup-2) and lower for sequential splittings (setup-3). In each panel, results from using different $\alpha_\mathrm{s}$ values are compared. For a given setup of formation time, the jet $R_\mathrm{AA}$ becomes smaller with an increasing value of $\alpha_\mathrm{s}$ due to stronger jet-medium interactions. Meanwhile, with the same $\alpha_\mathrm{s}$ value, an increasing $R_\mathrm{AA}$ can be observed from top to bottom panels, because a longer formation time ($t_\mathrm{init}^\mathrm{setup-3}>t_\mathrm{init}^\mathrm{setup-2}>t_\mathrm{init}^\mathrm{setup-1}$) delays the medium modification on jets. A similar trend of larger $R_\mathrm{AA}$ with later jet formation time was also observed in Ref.~\cite{Apolinario:2020uvt}. In addition to the overall magnitude of jet quenching, here we also notice the $p_\mathrm{T}$ dependence of the jet $R_\mathrm{AA}$ can be affected by different assumptions of the parton formation time. At the RHIC energy, the jet $R_\mathrm{AA}$ would decrease with $p_\mathrm{T}$ if instantaneous formation is assumed (upper panel). However, by adopting a more realistic modeling of formation time, a slightly rising trend of $R_\mathrm{AA}$ with respect to $p_\mathrm{T}$ is observed. This can be understood with the larger formation time for higher $p_\mathrm{T}$ partons, as previously discussed in Figs.~\ref{fig:ftvsFinalpT} and~\ref{fig:ftvsInitialpT}. This effect on the $p_\mathrm{T}$ dependence of the jet $R_\mathrm{AA}$ does not depend on the $\alpha_\mathrm{s}$ value we used here. Note that the STAR data~\cite{STAR:2020xiv} of $R_\mathrm{AA}^\mathrm{Pythia}$ (ratio of jet spectra between measurement in A+A collisions and Pythia simulation) are also included in Fig.~\ref{fig:Raa} as a reference. However, as discussed earlier, we did not aim to precisely constrain the formation time from data in this study yet, considering that our current model calculation is still incomplete. For instance, jet partons at high virtuality (before arriving at their mass shells) can also lose energy inside the QGP~\cite{JETSCAPE:2017eso,Cao:2021rpv}, which has not been taken into account in our current study. Therefore, results from our setup-3 in the lower panel of Fig.~\ref{fig:Raa} do not conclude the coupling strength should be as strong as $\alpha_\mathrm{s}=0.5$. In addition, due to the challenge remaining in hadronizing jet partons in heavy-ion collisions, uncertainties exist in comparing our partonic jets in the present study to the charged jets with a high $p_\mathrm{T}$ hadron trigger measured by STAR. These will be improved in our future efforts.

To avoid uncertainties introduced by lacking the p+p baseline of jet measurement, one may also quantify the nuclear modification effect using the central-to-peripheral ratio ($R_\mathrm{cp}$) of the jet spectra in A+A collisions. Shown in Fig.~\ref{fig:Rcp} is our calculation on this $R_\mathrm{cp}$ between 0-10\% and 60-80\% centrality bins, compared to the STAR data~\cite{STAR:2020xiv}. From the top to bottom panels, we present results for different jet cone sizes -- $R=0.2$, 0.3, and 0.4 respectively. In each panel, we compare between the three proposed setups of formation time estimation. Similar to the jet $R_\mathrm{AA}$ previously presented in Fig.~\ref{fig:Raa}, we found that different modelings of the parton formation time affect not only the overall magnitude, but also the $p_\mathrm{T}$ dependence of the jet $R_\mathrm{cp}$. Setup-1, that is, instantaneous parton formation, leads to a decreasing jet $R_\mathrm{cp}$ with respect to its $p_\mathrm{T}$, which is disfavored by the STAR data. By contrast, the increasing trend of $R_\mathrm{cp}$ with jet $p_\mathrm{T}$ from more realistic evaluations of the formation time (setup-2 and 3) appear qualitatively consistent with the experimental observation. Since formation time increases with the virtuality scale of the initial hard partons, its impact on jet quenching is expected to be even stronger in more energetic collisions at the LHC.

\section{Summary}
\label{sec:summary}

In this study, we explored the impact of the parton shower formation time on jet quenching in relativistic heavy-ion collisions. The Pythia event generator was used to obtain parton showers in vacuum, based on which three different models were set up to estimate the formation time of each parton, including instantaneous formation (setup-1), formation from single splitting (setup-2), and sequential splittings (setup-3). The final-state partons from Pythia were then fed into the LBT model for their subsequent interactions with the QGP medium.

Within this framework, we found that after considering the time taken by realistic splittings (for both setup-2 and 3), only a limited number of partons within jets form prior to the QGP formation, while a large number form inside and even after the QGP stage. The formation time becomes longer as the scale of the momentum exchange in the initial hard scatterings increases. Within the lifetime of QGP produced at  RHIC energy, we found that the average parton formation time follows the hierarchy $t_\mathrm{init}^\mathrm{setup-3}>t_\mathrm{init}^\mathrm{setup-2}>t_\mathrm{init}^\mathrm{setup-1}$, which leads to an inverse hierarchy of parton energy loss inside the QGP. Remarkable effects were observed on both the overall magnitude and transverse momentum dependence of the nuclear modification factor of jets. For a given value of $\alpha_\mathrm{s}$, a smaller $R_\mathrm{AA}$ was observed for a shorter formation time. The jet $R_\mathrm{AA}$ decreased with $p_\mathrm{T}$ for  setup-1, while increased with $p_\mathrm{T}$ for setup-2 and 3, owing to a longer average parton formation time from a more energetic jet. Consistent results were observed across different $\alpha_\mathrm{s}$ values, as well as for the central-to-peripheral ratio of the jet spectra ($R_\mathrm{cp}$).

While our study has provided a detailed demonstration on the sensitivity of jet quenching observables to the formation time of parton showers, it requires further improvements in several directions in order to achieve quantitative constraints on the parton formation time from the jet quenching data. For instance, instead of the free-streaming assumption, it is necessary to introduce a medium modification on jet partons during their high virtuality stage to avoid possible underestimation of jet quenching, especially when the parton formation time is long. This would narrow the difference between using various assumptions of the formation time, although different $p_\mathrm{T}$ dependences of jet observables are still expected due to different jet energy loss mechanisms at different virtuality scales. A solid hadronization scheme should also be introduced for a more direct comparison between our current model calculation at the parton level and experimental measurements on the charged hadron jets. Last but not least, more jet observables, such as the anisotropic flow coefficients of jets and the jet shape, should be included for a stronger constraint on the formation time.


\begin{acknowledgments} 
We thank Yayun He, Tan Luo, Weiyao Ke, Maowu Nie, and Xin-Nian Wang for helpful discussions. 
\end{acknowledgments}


\bibliographystyle{h-physrev5}
\bibliography{SCrefs}
\end{document}